\newcommand{\CLASSINPUTtoptextmargin}{19mm}%
\newcommand{\CLASSINPUTbottomtextmargin}{43mm}%
\newcommand{\CLASSINPUTinnersidemargin}{12.9mm}%
\newcommand{\CLASSINPUToutersidemargin}{12.9mm}%

\documentclass[conference,10pt,a4paper]{IEEEtran}
\usepackage{amsmath}
\usepackage{times}
\usepackage{graphicx}
\usepackage{multirow}
\usepackage[none]{hyphenat}
\usepackage{float}
\usepackage{subfig}

\usepackage{xfrac}
\usepackage[toc,nomain]{glossaries}
\newglossary[sym]{sym}{symin}{symout}{Symbolverzeichnis}

\DeclareMathOperator{\GHz}{\,GHz}

\usepackage{svg}

\usepackage{tikz}

\newcommand\copyrighttext{%
	\footnotesize \textcopyright 2023 IEEE. Personal use of this material is permitted.  Permission from IEEE must be obtained for all other uses, in any current or future media, including reprinting/republishing this material for advertising or promotional purposes, creating new collective works, for resale or redistribution to servers or lists, or reuse of any copyrighted component of this work in other works. DOI: 10.23919/EuRAD58043.2023.10289187}
\newcommand\copyrightnotice{%
	\begin{tikzpicture}[remember picture,overlay]
	\node[anchor=south,yshift=10pt] at (current page.south) {\fbox{\parbox{\dimexpr\textwidth-\fboxsep-\fboxrule\relax}{\copyrighttext}}};
	\end{tikzpicture}%
}

\newcommand{\VLaplace}[1][]{\mbox{\setlength{\unitlength}{0.1em}%
		\begin{picture}(10,20)%
		\put(3,2){\circle*{4}}%
		\put(3,4){\line(0,1){24}}%
		\put(3,30){\circle{4}}%
		\put(10,7){#1}
		\end{picture}%
	}%
}%

\newcommand{\Gl}[1]{(\ref{#1})}
\newcommand{\Abb}[1]{Fig.~\ref{#1}}

\makeatletter

\def\@maketitle{\newpage
\bgroup\par\addvspace{0.5\baselineskip}\centering%
\ifCLASSOPTIONtechnote
   {\bfseries\large\@IEEEcompsoconly{\sffamily}\@title\par}\vskip 1.3em{\lineskip .5em\@IEEEcompsoconly{\sffamily}\@author
   \@IEEEspecialpapernotice\par{\@IEEEcompsoconly{\vskip 1.5em\relax
   \@IEEEtitleabstractindextextbox{\@IEEEtitleabstractindextext}\par
   \hfill\@IEEEcompsocdiamondline\hfill\hbox{}\par}}}\relax
\else
   \vskip0.2em{\EuMWtitlesize\ifCLASSOPTIONtransmag\bfseries\LARGE\fi\@IEEEcompsoconly{\sffamily}\@IEEEcompsocconfonly{\normalfont\normalsize\vskip 2\@IEEEnormalsizeunitybaselineskip
   \bfseries\Large}\@title\par}\vskip1.0em\par
   \ifCLASSOPTIONconference%
      {\@IEEEspecialpapernotice\mbox{}\vskip\@IEEEauthorblockconfadjspace%
       \mbox{}\hfill\begin{@IEEEauthorhalign}\@author\end{@IEEEauthorhalign}\hfill\mbox{}\par}\relax
   \else
      \ifCLASSOPTIONpeerreviewca
         {\@IEEEcompsoconly{\sffamily}\@IEEEspecialpapernotice\mbox{}\vskip\@IEEEauthorblockconfadjspace%
          \mbox{}\hfill\begin{@IEEEauthorhalign}\@author\end{@IEEEauthorhalign}\hfill\mbox{}\par
          {\@IEEEcompsoconly{\vskip 1.5em\relax
           \@IEEEtitleabstractindextextbox{\@IEEEtitleabstractindextext}\par\hfill
           \@IEEEcompsocdiamondline\hfill\hbox{}\par}}}\relax
      \else
         \ifCLASSOPTIONtransmag
           {\@IEEEspecialpapernotice\mbox{}\vskip\@IEEEauthorblockconfadjspace%
            \mbox{}\hfill\begin{@IEEEauthorhalign}\@author\end{@IEEEauthorhalign}\hfill\mbox{}\par
           {\vspace{0.5\baselineskip}\relax\@IEEEtitleabstractindextextbox{\@IEEEtitleabstractindextext}\vspace{-1\baselineskip}\par}}\relax
         \else
           {\lineskip.5em\@IEEEcompsoconly{\sffamily}\sublargesize\@author\@IEEEspecialpapernotice\par
           {\@IEEEcompsoconly{\vskip 1.5em\relax
            \@IEEEtitleabstractindextextbox{\@IEEEtitleabstractindextext}\par\hfill
            \@IEEEcompsocdiamondline\hfill\hbox{}\par}}}\relax
         \fi
      \fi
   \fi
\fi\par\addvspace{0.0\baselineskip}\egroup}

\def\EuMWtitlesize{\@setfontsize{\EuMWtitlesize}{24}{24pt}}
\def\EuMWauthorsize{\@setfontsize{\EuMWauthorsize}{11}{11pt}}
\def\EuMWaffilsize{\@setfontsize{\EuMWaffilsize}{10}{10pt}}
\def\EuMWcaptionsize{\@setfontsize{\EuMWcaptionsize}{9}{10pt}}
\def\EuMWbibsize{\@setfontsize{\EuMWbibsize}{8}{10pt}}

\def\@IEEEauthorblockNstyle{\EuMWauthorsize\@IEEEcompsocnotconfonly{\sffamily}\@IEEEcompsocconfonly{\large}}
\def\@IEEEauthorblockAstyle{\EuMWaffilsize\@IEEEcompsocnotconfonly{\sffamily}\@IEEEcompsocconfonly{\itshape}\@IEEEcompsocconfonly{\large}}
\def\@IEEEauthordefaulttextstyle{\EuMWauthorsize\@IEEEcompsocnotconfonly{\sffamily}\sublargesize}

\def\thebibliography#1{\section*{\refname}%
    \addcontentsline{toc}{section}{\refname}%
    \EuMWbibsize\@IEEEcompsocconfonly{\small}\vskip 0.3\baselineskip plus 0.1\baselineskip minus 0.1\baselineskip
    \list{\@biblabel{\@arabic\c@enumiv}}%
    {\settowidth\labelwidth{\@biblabel{#1}}%
    \leftmargin\labelwidth
    \advance\leftmargin\labelsep\relax
    \itemsep \IEEEbibitemsep\relax
    \usecounter{enumiv}%
    \let\p@enumiv\@empty
    \renewcommand\theenumiv{\@arabic\c@enumiv}}%
    \let\@IEEElatexbibitem\bibitem%
    \def\bibitem{\@IEEEbibitemprefix\@IEEElatexbibitem}%
\def\newblock{\hskip .11em plus .33em minus .07em}%
\ifCLASSOPTIONtechnote\sloppy\clubpenalty4000\widowpenalty4000\interlinepenalty100%
\else\sloppy\clubpenalty4000\widowpenalty4000\interlinepenalty500\fi%
    \sfcode`\.=1000\relax}

\long\def\@makecaption#1#2{%
\ifx\@captype\@IEEEtablestring%
\par\@IEEEtabletopskipstrut
\else
\@IEEEfigurecaptionsepspace
\fi
\setbox\@tempboxa\hbox{\normalfont\footnotesize {#1.}\nobreakspace\nobreakspace #2}%
\ifdim \wd\@tempboxa >\hsize%
\setbox\@tempboxa\hbox{\normalfont\footnotesize {#1.}\nobreakspace\nobreakspace}%
\parbox[t]{\hsize}{\normalfont\footnotesize\noindent\unhbox\@tempboxa#2}%
\else
\ifCLASSOPTIONconference \hbox to\hsize{\normalfont\footnotesize\hfil\box\@tempboxa\hfil}%
\else \hbox to\hsize{\normalfont\footnotesize\box\@tempboxa\hfil}%
\fi\fi
\ifx\@captype\@IEEEtablestring%
\@IEEEtablecaptionsepspace
\else
\fi}

\newlength\tablecaptiontotableskip
\newlength\figuretocaptionskip
\def\@IEEEfigurecaptionsepspace{\vskip\figuretocaptionskip\relax}%
\def\@IEEEtablecaptionsepspace{\vskip\tablecaptiontotableskip\relax}%

\def\abstract{\normalfont%
\@IEEEabskeysecsize\bfseries\textit{\abstractname}\,\bfseries\textit{---}\,%
\@IEEEgobbleleadPARNLSP}%

\def\IEEEkeywords{\normalfont%
\@IEEEabskeysecsize\bfseries\textit{\IEEEkeywordsname}\,\bfseries\textit{---}\,%
\@IEEEgobbleleadPARNLSP}%
\def\endIEEEkeywords{\relax\vspace{0.67ex}%
\par\if@twocolumn\else\endquotation\fi%
\normalsize\normalfont}%

\DeclareRobustCommand*{\EuMWauthorrefmark}[1]{\raisebox{0pt}[0pt][0pt]{\textsuperscript{\footnotesize{#1}}}}%
\def\@IEEEauthorblockNtopspace{0ex}
\def\@IEEEauthorblockAtopspace{1mm}
\def\tablename{Table}
\def\thetable{\arabic{table}}
\def\IEEEkeywordsname{Keywords}
\def\subsubsection{\@startsection{subsubsection}{3}{\z@}{1.5ex plus 1.5ex minus 0.5ex}%
{0.7ex plus .5ex minus 0ex}{\normalfont\normalsize\itshape}}%
\newlength{\CPheadmatchindent}%
\def\@seccntformat#1{\hbox to\CPheadmatchindent{\csname the#1dis\endcsname}\hskip 0.1em \relax}
\IEEEilabelindentA \parindent
\IEEEilabelindent \IEEEilabelindentA
\IEEEelabelindent \parindent
\IEEEdlabelindent \parindent
\IEEElabelindent \parindent
\makeatother

\newglossaryentry{fc}{
	type=sym,
	name={\ensuremath{f_{\mathrm{c}}}},
	symbol=\ensuremath{Hz},
	sort=fc,
	description={Mittenfrequenz des Radars}}

\newglossaryentry{f0}{
	type=sym,
	name={\ensuremath{f_{\mathrm{c}}}},
	symbol=\ensuremath{Hz},
	sort=f0,
	description={Mittenfrequenz des Radars}}

\newglossaryentry{hid}{
	type=sym,
	name={\ensuremath{h_{\mathrm{id}}}},
	symbol=\ensuremath{},
	sort=hid,
	description={Idealisierte Kanalimpulsantwort}}

\newglossaryentry{hest}{
	type=sym,
	name={\ensuremath{\tilde{h}}},
	symbol=\ensuremath{},
	sort=htilde,
	description={Geschätzte Kanalimpulsantwort}}

\newglossaryentry{Hid}{
	type=sym,
	name={\ensuremath{H_{\mathrm{id}}}},
	symbol=\ensuremath{},
	sort=Hid,
	description={Idealisierte Kanalübertragungsfunktion}}

\newglossaryentry{speed}{
	type=sym,
	name={\ensuremath{\nu}},
	symbol=\ensuremath{m/s},
	sort=Hid,
	description={Geschwindigkeit des Ziels}}

\newglossaryentry{wavelength}{
	type=sym,
	name={\ensuremath{\lambda}},
	symbol=\ensuremath{m},
	sort=Lambda,
	description={Wellenlänge}}

\newglossaryentry{distance}{
	type=sym,
	name={\ensuremath{R}},
	symbol=\ensuremath{m},
	sort=R,
	description={Distanz zum Ziel}}

\newglossaryentry{angle}{
	type=sym,
	name={\ensuremath{\alpha}},
	symbol=\ensuremath{rad},
	sort=alpha,
	description={Winkel zum Ziel}}

\newglossaryentry{rampnumber}{
	type=sym,
	name={\ensuremath{m}},
	symbol=\ensuremath{},
	sort=m,
	description={Nummer der Rampe im Messzyklus}}

\newglossaryentry{numberoframps}{
	type=sym,
	name={\ensuremath{M}},
	symbol=\ensuremath{},
	sort=M,
	description={Anzahl der Frequenzrampen in einem Messzyklus}}

\newglossaryentry{antennanumber}{
	type=sym,
	name={\ensuremath{n}},
	symbol=\ensuremath{},
	sort=n,
	description={Nummer der Antenne}}

\newglossaryentry{numberofantennas}{
	type=sym,
	name={\ensuremath{N}},
	symbol=\ensuremath{},
	sort=N,
	description={Anzahl der Antennen}}

\newglossaryentry{antennadistance}{
	type=sym,
	name={\ensuremath{u}},
	symbol=\ensuremath{},
	sort=u,
	description={Distanz zwischen den Antennen}}

\newglossaryentry{sampleinterval}{
	type=sym,
	name={\ensuremath{T_A}},
	symbol=\ensuremath{},
	sort=TA,
	description={Abtastinterval}}

\newglossaryentry{samplerate}{
	type=sym,
	name={\ensuremath{F_A}},
	symbol=\ensuremath{},
	sort=FA,
	description={Abtastrate}}

\newglossaryentry{samplenumber}{
	type=sym,
	name={\ensuremath{p}},
	symbol=\ensuremath{},
	sort=p,
	description={Nummer des Abtastpunkts im Zeitbereich}}

\newglossaryentry{numberofsamples}{
	type=sym,
	name={\ensuremath{P}},
	symbol=\ensuremath{},
	sort=P,
	description={Anzahl der Abtastpunkte im Zeitbereich}}

\newglossaryentry{ft}{
	type=sym,
	name={\ensuremath{f(t)}},
	symbol=\ensuremath{\mathrm{Hz}},
	sort=f,
	description={Momentanfrequenz}}

\newglossaryentry{stx}{
	type=sym,
	name={\ensuremath{s^{tx}(t)}},
	symbol=\ensuremath{\mathrm{V}},
	sort=f,
	description={Generatorsignal}}

\newglossaryentry{srx}{
	type=sym,
	name={\ensuremath{s^{}rx}(t)},
	symbol=\ensuremath{\mathrm{V}},
	sort=f,
	description={Empfangendes Signal}}

\newglossaryentry{deltaf}{
	type=sym,
	name={\ensuremath{\Delta f}},
	symbol=\ensuremath{Hz},
	sort=fdelta,
	description={Bandbreite}}

\newglossaryentry{transmitpower}{
	type=sym,
	name={\ensuremath{X_0}},
	symbol=\ensuremath{W},
	sort=X0,
	description={Sendeleistung}}

\newglossaryentry{Fp}{
	type=sym,
	name={\ensuremath{F_{\mathrm{p}}}},
	symbol=\ensuremath{1},
	sort=Fp,
	description={Ausbreitungsfaktor}}

\newglossaryentry{sweepslope}{
	type=sym,
	name={\ensuremath{\alpha}},
	symbol=\ensuremath{Hz/s},
	sort=alpha2,
	description={Steilheit der Frequenzrampe}}

\newglossaryentry{Tsw}{
	type=sym,
	name={\ensuremath{T_{sw}}},
	symbol=\ensuremath{s},
	sort=Tsw,
	description={Rampendauer}}

\newglossaryentry{Tpri}{
	type=sym,
	name={\ensuremath{T_{\mathrm{PRI}}}},
	symbol=\ensuremath{s},
	sort=Tpri,
	description={Rampenwiederholinterval, engl. Pulse Repitition Interval}}

\newglossaryentry{Tcpi}{
	type=sym,
	name={\ensuremath{T_{\mathrm{CPI}}}},
	symbol=\ensuremath{s},
	sort=Tpri,
	description={Kohärentes Verarbeitungsinterval, engl. Coherent Processing Interval}}

\newglossaryentry{numberoftargets}{
	type=sym,
	name={\ensuremath{N_T}},
	symbol=\ensuremath{},
	sort=NT,
	description={Anzahl der Detektionen}}

\newglossaryentry{tau}{
	type=sym,
	name={\ensuremath{\tau}},
	symbol=\ensuremath{s},
	sort=tau,
	description={Umlaufzeit}}

\newglossaryentry{c0}{
	type=sym,
	name={\ensuremath{c_0}},
	symbol=\ensuremath{m/s},
	sort=c0,
	description={Lichtgeschwindigkeit}}

\newglossaryentry{tmess}{
	type=sym,
	name={\ensuremath{t}},
	symbol=\ensuremath{s},
	sort=tmess,
	description={Zeit}}

\newglossaryentry{height_resolution}{
	type=sym,
	name={\ensuremath{\delta h}},
	symbol=\ensuremath{m},
	sort=deltah,
	description={Höhenauflösung}}

\newglossaryentry{elevation_angle}{
	type=sym,
	name={\ensuremath{\beta}},
	symbol=\ensuremath{\deg},
	sort=elevationangle,
	description={Elevationswinkel}}

\newglossaryentry{target_height}{
	type=sym,
	name={\ensuremath{h_{\mathrm{T}}}},
	symbol=\ensuremath{m},
	sort=deltah,
	description={Höhe des Ziels}}

\newglossaryentry{sensor_height}{
	type=sym,
	name={\ensuremath{h_{\mathrm{S}}}},
	symbol=\ensuremath{m},
	sort=deltah,
	description={Höhe des Sensors}}

\newglossaryentry{reflection_factor}{
	type=sym,
	name={\ensuremath{\Gamma}},
	symbol=\ensuremath{1},
	sort=gamma,
	description={Bodenreflexionsfaktor}}

\newglossaryentry{reflection_factor_linse}{
	type=sym,
	name={\ensuremath{\gls{reflection_factor}_{Linsenfaktor}}},
	symbol=\ensuremath{1},
	sort=gammaLinse,
	description={Divergenzfaktor}}

\newglossaryentry{reflection_factor_fresnel}{
	type=sym,
	name={\ensuremath{\gls{reflection_factor}_{Fresnel}}},
	symbol=\ensuremath{1},
	sort=gammaFresnel,
	description={Fresnel-Reflexionskoeffizient}}

\newglossaryentry{reflection_factor_rau}{
	type=sym,
	name={\ensuremath{\gls{reflection_factor}_{Rauheit}}},
	symbol=\ensuremath{1},
	sort=gammarau,
	description={Dämpfung des Bodenreflexionsfaktors durch Rauheit }}

\newglossaryentry{propfac}{
	type=sym,
	name={\ensuremath{F_{\mathrm{p}}}},
	symbol=\ensuremath{1},
	sort=deltah,
	description={Propagation Faktor}}

\newglossaryentry{distance_over_ground}{
	type=sym,
	name={\ensuremath{d}},
	symbol=\ensuremath{m},
	sort=deltah,
	description={Horizontalabstand zwischen Sensor und Ziel}}

\newglossaryentry{height_spectrum}{
	type=sym,
	name={\ensuremath{H_{\mathrm{T}}}},
	symbol=\ensuremath{},
	sort=deltah,
	description={Höhenspektrum (Frequenzspektrum des Amplitudenfaktors)}}

\newglossaryentry{distance_sample_number}{
	type=sym,
	name={\ensuremath{n}},
	symbol=\ensuremath{},
	sort=deltah,
	description={Nummer des Distanzabtastpunktes}}

\newglossaryentry{std_h}{
	type=sym,
	name={\ensuremath{\sigma_{h}}},
	symbol=\ensuremath{},
	sort=sigmah,
	description={Standardabweichung der Höhe der Oberfläche zur Modellierung der Oberflächenrauheit}}

\newglossaryentry{std_h_mic}{
	type=sym,
	name={\ensuremath{\sigma_{h, mic}}},
	symbol=\ensuremath{\mathrm{m}},
	sort=sigmahmic,
	description={Standardabweichung der Höhe der Oberfläche zur Modellierung der Oberflächenfeinrauheit}}

\newglossaryentry{std_h_mac}{
	type=sym,
	name={\ensuremath{\sigma_{h, mac}}},
	symbol=\ensuremath{\mathrm{m}},
	sort=sigmahmac,
	description={Standardabweichung der Höhe der Oberfläche zur Modellierung der Oberflächenunebenheit}}

\newglossaryentry{std_phi}{
	type=sym,
	name={\ensuremath{\sigma_{\varphi}}},
	symbol=\ensuremath{\rad},
	sort=sigmaphi,
	description={Standardabweichung der Phasenunterschiede der an einer rauen Oberfläche reflektierten Wellenfront}}

\newglossaryentry{delta_phi}{
	type=sym,
	name={\ensuremath{\Delta \varphi}},
	symbol=\ensuremath{\rad},
	sort=deltaphi,
	description={Phasenunterschied der an einer rauen Oberfläche reflektierten Wellenfront}}

\newglossaryentry{delta_r}{
	type=sym,
	name={\ensuremath{\Delta r}},
	symbol=\ensuremath{\mathrm{m}},
	sort=deltar,
	description={Weglängenunterschied der an einer rauen Oberfläche reflektierten Wellenfront}}

\newglossaryentry{delta_h}{
	type=sym,
	name={\ensuremath{\Delta h}},
	symbol=\ensuremath{\mathrm{m}},
	sort=deltah,
	description={Abweichung der Höhe der Oberfläche zur Modellierung der Oberflächenrauheit}}

\newglossaryentry{erwartungswert}{
	type=sym,
	name={\ensuremath{E}},
	symbol=\ensuremath{},
	sort=erwartungswert,
	description={Erwartungswert}}

\newglossaryentry{length_fresnel_zone}{
	type=sym,
	name={\ensuremath{l_F}},
	symbol=\ensuremath{m},
	sort=lengthfresnelzone,
	description={Länge der Fresnel Zone der Bodenreflexion}}

\newglossaryentry{unebenheitsmass}{
	type=sym,
	name={\ensuremath{\Phi_0}},
	symbol=\ensuremath{cm$^3$},
	sort=phi0,
	description={Unebenheitsmaß}}

\newglossaryentry{bezugskreisfrequenz}{
	type=sym,
	name={\ensuremath{\Omega_0}},
	symbol=\ensuremath{1\,m$^{-1}$},
	sort=omega0,
	description={Bezugswert für die Wegkreisfrequenz}}

\begin{document}
\raggedbottom
%
%
%
\title{FMCW Radar Height Estimation of Moving Vehicles by Analyzing Multipath Reflections}
%
%
\author{%
\IEEEauthorblockN{%
S\"oren Kohnert\EuMWauthorrefmark{\#1}, 
Michael Vogt\EuMWauthorrefmark{*2}, 
Reinhard Stolle\EuMWauthorrefmark{\#3}
}
\IEEEauthorblockA{%
\EuMWauthorrefmark{\#}HSA\_ired, Augsburg Technical University of Applied Sciences, 86161 Augsburg, Germany\\
\EuMWauthorrefmark{*}Institute of Electronic Circuits, Ruhr University Bochum, 44801 Bochum, Germany\\
\{\EuMWauthorrefmark{1}soeren.kohnert, \EuMWauthorrefmark{3}reinhard.stolle\}@hs-augsburg.de, \EuMWauthorrefmark{2}michael.vogt@rub.de\\
}
}
%
\maketitle
\copyrightnotice
%
%
\begin{abstract}
Target classification is an important task of automotive radar systems.
In this work, a concept for estimating the height of vehicles to allow for a differentiation between passenger cars, trucks, and others, is presented and discussed.
Fixed installed radar sensors for traffic monitoring in the 77\,GHz band are used to track and analyze radar echoes from individual vehicles as they move relative to the radar.
Considering multipath propagation, which includes the ground reflection, the height of individual radar targets is estimated by analyzing the periodicity of the resulting amplitude modulation (AM) of the echo signal as a function of the horizontal distance from the radar.
Two approaches have been realized to integrate the concept into an automotive FMCW signal processing scheme.
%
Measurements in a test field using a trihedral corner reflector as an idealized target in heights from 0 to 2.5\,m suggest, that the concept is suitable for height estimation in an automotive context.
\end{abstract}
%
%

\section{Introduction}
%
In the context of automated driving, the class of vehicles can give hints about their expected behavior (e.g., driven velocity) and their characteristics (e.g., length). 
Among other sensors, radar sensors provide information for a corresponding classification of vehicles by extracting suitable features \cite{visentin_2017}, such as the radar cross section (RCS).
In this work a concept for estimating the height of vehicles is investigated that bases on the multipath propagation between the radar and the vehicle. Estimated heights can be used as a feature in a possible subsequent target classification system.
The wave transmitted by the radar propagates both, the direct line of side and an additional path over the ground reflection, before and after the reflection at the vehicle: a total of four different paths between transmission and reception.
The superposition of the four echoes, each with a different time of flight, leads to amplitude modulation (AM) of the radar echo as a function of the horizontal distance between the sensor and the vehicle.
The frequency of this AM as a function of the reciprocal distance is proportional to the target height, and the corresponding information is employed here for height estimation.
%
%
%

In an automotive context, e.g., Diewald et al. \cite{diewald_2011_bridge} used a feature based on this effect to differentiate between bridges and stationary vehicles.
We have previously shown that it is possible to derive a classification feature by using a similar concept to distinguish between cars and trucks \cite{kohnert_2022_height}.

This contribution shows quantitatively how well a Fourier transform can be used to perform height estimation by analyzing the AM of the radar echo. The proposed concept has been implemented in a typical frequency modulated continuous wave (FMCW) radar signal processing scheme and is experimentally verified using a trihedral corner reflector as an idealized radar target. Limits of the concept are derived analytically: the smallest and largest distance in which it is applicable and the achievable resolution with the height estimation.
\section{Multipath propagation}
\vspace{-0.2cm}
    \begin{figure}[H]
	\centering
	\includesvg[width=\columnwidth]{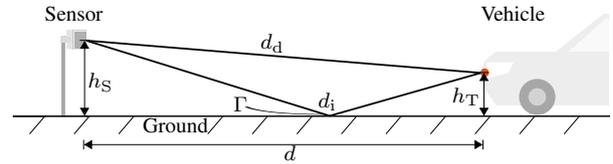}
	\caption{
		Geometry of the multipath model. On the left
		side is the stationary radar sensor at height $\gls{sensor_height}$. In the
		distance above ground $\gls{distance_over_ground}$, the scattering center (red dot) is located at height $\gls{target_height}$.}
	\label{fig:feat:hoehe:signalModel}
\end{figure}
\vspace{-0.3cm}
\Abb{fig:feat:hoehe:signalModel} shows the multipath model based on a car approaching the sensor.
%
The indirect path over the ground reflection only exists for spherical targets, targets with strong curvature, or targets scattering diffusely. B\"uhren et al. \cite{buehren_2007} show that the indirect path is present in the radar echo of cars. 
The reflections on the front and back of the car are thus approximated as point targets \cite{diewald_2011_bridge} i.e., isotropic scattering is assumed. 
\gls{target_height} may vary, but is on average lower for cars than for trucks.
%
%
The transfer function of the radar path
 \begin{align}
 H&(f)=  A\cdot  \exp\left({-j2\pi f\gls{tau}_{\mathrm{d}}}\right) \label{uebertragungsfunktionMitFp} \\ 
 \cdot & \underbrace{\big( 1 +  2\cdot \exp\left({-j2\pi f \Delta \gls{tau}}\right)\cdot \Gamma +  \exp\left({-j2\pi f 2 \Delta \gls{tau}}\right)\cdot \Gamma^2 \big) }_{\gls{propfac}} \nonumber
 \end{align}
  is obtained from the addition of all components of the four propagation paths.
  The modulation of the radar echo caused by the multipath propagation is described by the propagation factor \gls{propfac}.
  $\gls{reflection_factor}$ is the ground reflection coefficient. $\gls{tau}_{\mathrm{d}}$ is the direct path (or line of sight) propagation delay. $f$ is the center frequency of the radar. 
The attenuation $A$ and reflection coefficient of the target are approximated as equal for all paths. The propagation delay difference between the indirect and direct paths is $\Delta \gls{tau} = \gls{tau}_{\mathrm{i}} -  \gls{tau}_{\mathrm{d}}$.
A metallic plate is a suitable model for the ground in this application: for flat angles of incidence and smooth road surfaces the reflection coefficient tends towards $\Gamma = -1$. If $|\Gamma| < 1$, a direct current (DC) component is superimposed on the propagation factor.
With that, the propagation factor simplifies to
\begin{equation}
\begin{split}
\gls{propfac}(f) = -4\cdot \exp(-j2\pi f\Delta\tau) \cdot \sin^2(2\pi f \Delta \gls{tau} / 2) 
\end{split}
\label{eq:feat:hoehe:Fp3}
\end{equation}
An approximation for the difference of the propagation delay is derived from the geometry in \Abb{fig:feat:hoehe:signalModel} and a series expansion to
\vspace{-0.1cm}
\begin{equation}
\Delta \gls{tau} \approx \frac{2\gls{target_height} \gls{sensor_height}}{c \cdot d},
\label{eq:approxDiffDistances}
\end{equation}
where $c$ is the speed of light. The approximation is good if $\gls{distance_over_ground}^2 \gg \gls{sensor_height}^2,\gls{target_height}^2$. 
Thus, the magnitude of the propagation factor becomes
\vspace{-0.2cm}
\begin{equation}
|\gls{propfac}(\sfrac{1}{\gls{distance_over_ground}})| \approx 4 \cdot \sin^2\left(2\pi \frac{\gls{sensor_height}}{\gls{wavelength}}\cdot \gls{target_height} \cdot \sfrac{1}{d} \right),
\label{eq:feat:hoehe:betrag_Fp_d}
\end{equation}
where \gls{wavelength} is the wavelength. In order to allow for a superposition of the radar echoes traveling along the different paths, the maxima of the impulse responses of the individual echo signals have to overlap inside one range resolution cell of the radar:
\begin{equation}
{}_\Delta  \gls{distance} \overset{!}{>} \frac{4\gls{target_height} \gls{sensor_height}}{d},
\label{eq::feat:hoehe:minDeltaR}
\end{equation}
where ${}_\Delta  \gls{distance}$ is the radar range resolution.
This defines a lower bound distance for the AM of the resulting radar echo to be described by \Gl{eq:feat:hoehe:betrag_Fp_d}.
%
Several scattering centers in a common range resolution cell of the same strength result in an arbitrarily complicated expression due to the absolute value formation in \Gl{eq:feat:hoehe:betrag_Fp_d}. Thus, one scattering center on the vehicle must be dominant, otherwise \Gl{eq:feat:hoehe:betrag_Fp_d} does not apply. %

\section{Height Estimation}
The parameters in the argument of \Gl{eq:feat:hoehe:betrag_Fp_d} are known a priori (wavelength \gls{wavelength}, sensor height \gls{sensor_height}) or can be approximated from the impulse response of the measurement (distance \gls{distance_over_ground}), only the target height \gls{target_height} is yet to be found.
\vspace{-0.1cm}
\begin{figure}[h]
	\centering
	\includegraphics[trim={0.1cm 6.1cm 1cm 0.3cm},clip,width=\columnwidth]{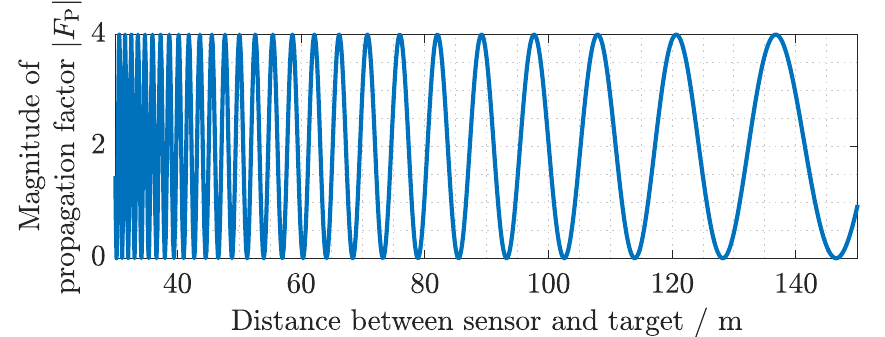}
	\caption{
		Magnitude of the propagation factor plotted versus distance between the sensor and the target for a height of sensor and target of $1$\,m, a center frequency of $76.5\GHz$, and a ground reflection coefficient of $\gls{reflection_factor}=-1$.
	}
	\label{fig:feat:hoehe:propfac}
\end{figure}
\vspace{-0.2cm}
The AM frequency of the propagation factor as a function of the reciprocal distance $|\gls{propfac}(\sfrac{1}{\gls{distance_over_ground}})|$ is proportional to the target height \gls{target_height}. In the use case of an approaching vehicle, the distance between the car and the sensor changes over time.
\Abb{fig:feat:hoehe:propfac} shows an example of the variation of the propagation factor plotted against the distance between the sensor and the target.
The Fourier transform of the magnitude of the propagation factor as a function of the reciprocal distance \Gl{eq:feat:hoehe:betrag_Fp_d} is
%
\begin{equation}
\begin{split}
 \gls{height_spectrum}(h) \approx 2 \cdot \delta(h) -  \delta\bigg(h-\underbrace{\frac{2\gls{sensor_height}}{\gls{wavelength}}\cdot \gls{target_height}}_{h_0} \bigg) - \delta\bigg(h+\underbrace{\frac{2\gls{sensor_height}}{\gls{wavelength}}\cdot \gls{target_height}}_{h_0} \bigg). \\
\end{split}
\label{eq:feat:hoehe:hoehenspektrum}
\end{equation}
\Gl{eq:feat:hoehe:hoehenspektrum} has a DC component and spectral components at a frequency proportional to the target height:
\begin{equation}
\gls{target_height} = \frac{\gls{wavelength}}{2 \gls{sensor_height}} h_0 .
\label{eq::feat:hoehe:umrechnung}
\end{equation}
The location of the maximum in the spectrum allows an estimate of the target height \gls{target_height}.
\subsection{Spectral Resolution}
\label{resolution}
The propagation factor is observed over a changing distance between the vehicle and the radar. The according distance interval $\Delta d$ in the reciprocal distance domain is inversely proportional to the achievable spectral resolution $\Delta {\gls{target_height}}$. 
Under the assumptions of equal target amplitudes, rectangular windowing, and a possible target separation up to half the main lobe width of the sinc pulses in the spectrum, the spectral resolution is 
\begin{equation}
\begin{split}
\Delta \gls{target_height} = & \frac{\gls{wavelength}\left(d_0^2 - \Delta d^2/4\right) }{2 \gls{sensor_height}  \Delta d},\\
\end{split}
\label{eq:feat:hoehe:delta_ht}
\end{equation}
where $d_0$ is the distance between sensor and target at the center of the observation interval.
The spectral resolution is proportional to the achievable accuracy and determines the minimum height that can be estimated at a given distance.
The minimum distance to be evaluated in order to achieve a desired resolution in the spectrum can be derived from \Gl{eq:feat:hoehe:delta_ht}:
\begin{equation}
\begin{split}
\Delta \gls{distance_over_ground} = 2 \cdot \left( \frac{-2\gls{sensor_height}\Delta \gls{target_height} + \sqrt{( 2\gls{sensor_height} \Delta \gls{target_height} )^2 + (\gls{wavelength}d_0)^2}}{\gls{wavelength}} \right).
\end{split}
\label{eq:feat:hoehe:delta_d}
\end{equation}
\subsection{Frequency Estimation}
All frequency estimation methods that can handle non-equidistantly sampled values can be considered. In the following, two possibilities are presented.

\subsubsection{Non-equidistant DFT}
\label{chap:feat:hoehe:abtastung:näquiDFT}
For the estimation of the spectrum \Gl{eq:feat:hoehe:hoehenspektrum}, a DFT is used:
\begin{equation}
\hat{H}_{\mathrm{T}}[k]  = W \cdot |\gls{propfac}(\gls{distance_over_ground} [\gls{distance_sample_number}])| \,\,\,\,\,\, \textnormal{ with } \,\,\,\,\,\, W[k,\gls{distance_sample_number}] = e^{-2j \pi \cdot \gls{distance_over_ground} [\gls{distance_sample_number}] \cdot k\Delta_{\mathrm{h}} }.
\label{chap:feat:hoehe:dft:matrix}
\raisetag{-1cm}
\end{equation}
$\Delta_{\mathrm{h}}$  is the distance between the support points in the spectrum $\hat{H}_{\mathrm{T}}[k]$.
The spectrum $\hat{H}_{\mathrm{T}}[k]$ has to be calculated only for the heights relevant to the application. The negative frequencies of the spectrum $\hat{H}_{\mathrm{T}}[k]$ can also be neglected due to the symmetry of real signals in the frequency domain.
Analogous to the scaling of the height axis for the continuous spectrum $H_{\mathrm{T}}(h)$ \Gl{eq::feat:hoehe:umrechnung}, the height axis can be scaled in the discrete case utilizing
\begin{equation}
\Delta_{\gls{target_height}} = \frac{ \gls{wavelength} }{2 \gls{sensor_height}} \Delta_{\mathrm{h}}.
\label{eq:feat:hoehe:dft:skalierung}
\end{equation}
\vspace{-0.4cm}
\subsubsection{Linearization}
\label{chap:feat:hoehe:abtastung:linDFT}
For large distances between the sensor and target in relation to a small observation interval $|\Delta \gls{distance_over_ground}| \ll \gls{distance_over_ground}_0$, the problem of non-equidistant sampling can be circumvented.
By a series expansion around the point $\gls{distance_over_ground}_0$,
\begin{equation}
 \frac{1}{\gls{distance_over_ground}} \approx \frac{1}{\gls{distance_over_ground}_0} \cdot \left( 1+\frac{\Delta \gls{distance_over_ground}}{\gls{distance_over_ground}_0}\right)
\label{eq:feat:hoehe:d_approx}
\end{equation}
can be approximated. 
Thus, the propagation factor from \Gl{eq:feat:hoehe:betrag_Fp_d} can be approximated to yield the corresponding Fourier transform pair
\vspace{-0.3cm}
\begin{align}
& |\gls{propfac}(\Delta \gls{distance_over_ground})| \approx 4 \cdot \sin^2\left(2\pi \frac{\gls{sensor_height}}{\gls{wavelength}}\cdot \gls{target_height} \cdot \frac{1}{\gls{distance_over_ground}_0} \left(1+\frac{\Delta \gls{distance_over_ground}}{\gls{distance_over_ground}_0} \right) \right) \nonumber \\
\VLaplace & H_{\mathrm{T}}(h) \approx 2 \cdot \delta(h) -  \delta\bigg(h-\frac{2\gls{sensor_height}}{\gls{wavelength}\gls{distance_over_ground}_0^2}\cdot \gls{target_height} \bigg) \cdot \exp(j2\pi \frac{2\gls{sensor_height}}{\gls{wavelength}\gls{distance_over_ground}_0}\cdot \gls{target_height}) \nonumber \\
& \hspace{1cm} - \delta\bigg(h+\underbrace{\frac{2\gls{sensor_height}}{\gls{wavelength}\gls{distance_over_ground}_0^2}\cdot \gls{target_height}}_{h'_0} \bigg)
\cdot \exp(-j2\pi \frac{2\gls{sensor_height}}{\gls{wavelength}\gls{distance_over_ground}_0}\cdot \gls{target_height}).
\label{eq:feat:hoehe:betrag_Fp_d_approx}
\end{align}
The height axes can be scaled using
\begin{equation}
\gls{target_height} = \frac{ \gls{wavelength} \gls{distance_over_ground}_0^2 }{2 \gls{sensor_height}} h'_0
\label{eq:feat:hoehe:lin_dft:skalierung}
\end{equation}
to the target height.
The approximation of the propagation factor can be considered periodic over the distance around the linearization point $\gls{distance_over_ground}_0$. Consequently, for the transformation of the linearized propagation factor \Gl{eq:feat:hoehe:betrag_Fp_d_approx}, a FFT can be used, which reduces the computational effort.

\section{FMCW Signal Processing}
The signal processing builds on the typical waveform of an automotive FMCW radar \cite{pourvoyeur_2008}.
A coherent processing interval (CPI) consists of a given number \gls{numberoframps} of successive frequency sweeps, which are used for range-Doppler measurement. First, a Fourier transform in range direction is calculated, yielding the impulse response from the channel transfer function. Next, a Fourier transform over the acquired sequence is calculated at each range gate, yielding the corresponding Doppler spectra. The time interval between two consecutive CPIs is available for signal processing.

%
\subsubsection{Sampling over the ramp repetition interval (S.\,o.\,PRI.)}

With this first realization, one sample of the target's AM is obtained from each individual impulse response measurement in the CPI.
That corresponds to a sampling interval equal to the ramp repetition interval, i.e., $\gls{Tpri} = \gls{Tcpi}/\gls{numberoframps}$, with \gls{numberoframps} being the number of ramps.
Typically, the sampling rate with S.\,o.\,PRI. ranges from 10 to 20\,kHz \cite{winner:2015}.
The range resolution of an FMCW radar is typically insufficient to determine the change in the position of a target between impulse response measurements based on a shift in the maximum magnitude.
For S.\,o.\,PRI., it is sufficient to determine only the relative position (as for Doppler estimation) and combine this with the (less accurate) position estimate by analyzing the maximum in the impulse responses.
The relative target position is calculated from the Doppler estimation by assuming a constant velocity.
The advantage of S.\,o.\,PRI. is the high sampling rate.
Disadvantageously, separating different radar targets from each other can only be performed by using range information but not Doppler information. 
The target's velocity defines the observed distance of the target if the observation is limited in duration, e.g., one CPI. The measurements in Section \ref{meas:PRI} show that if the AM is sampled only during one CPI with parameters typical for automotive radar, the observation interval limits the spectral resolution and, thus, the height estimation capabilities.
\subsubsection{Sampling over the measurement cycles (S.\,o.\,MC.)}
With this second realization, one sample of the target's AM is obtained from each target in the range-Doppler map. That corresponds to a sampling interval equal to the measurement cycle $T_{\mathrm{MC}}$. S.\,o.\,MC. allows for better separation of targets since the Doppler information is incorporated.
Disadvantageously, the sampling rate of S.\,o.\,MC. is lower (typically ranges from 9 to 20 Hz \cite{winner:2015}) compared to S. o. PRI.
%
\section{Experimental Evaluation}
\label{sec:measurements}
The processing to extract the AM of the target is realized utilizing a target extraction on the range-Doppler map followed by multi-target tracking over time \cite{pourvoyeur_2008}.
\vspace{-0.2cm}
\begin{figure}[H]
\centering
	\includesvg[width=10cm]{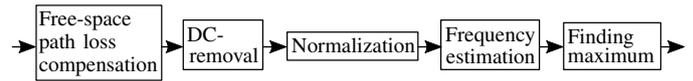}
	\caption{Signal-flow graph of the height estimation algorithm.}
	\label{fig:feat:hoehe:test:erg:ablaufplan}
\end{figure}
\vspace{-0.2cm}
\Abb{fig:feat:hoehe:test:erg:ablaufplan} shows the signal-flow graph of the height estimation algorithm. The input data are distances and the amplitudes of the echoes for each target. The free space attenuation is corrected by the factor of the squared distance between the sensor and target $\gls{distance_over_ground}^{2}$. The DC component is removed in the second step.
To make the AM spectrum independent of the reflection factor of the ground and independent of the RCS, the propagation factor is normalized to its maximum.
The DFT for non-equidistant sampled input data calculates a power spectral density (PSD). The estimated height can be read off the position of the maximum of the PSD.
\subsection{Measurement Setup}
\label{sec:meas}

For the measurements presented in this paper, different from the targeted application and due to a less complex practical implementation,
a stationary trihedral corner reflector (\Abb{fig:feat:hoehe:verif:SoRD:aufbau}~b))  and a radar system mounted in the rear of a car (\Abb{fig:feat:hoehe:verif:SoRD:aufbau}~a)) have been used.
\vspace{-0.2cm}
\begin{figure}[h]
	\raggedright
	\hspace{0.1cm}
	\begin{minipage}{0.48\columnwidth}
		\raggedleft
		\includegraphics[trim={2.35cm 1.6cm 1cm 2cm},clip,angle=0, height = \columnwidth]{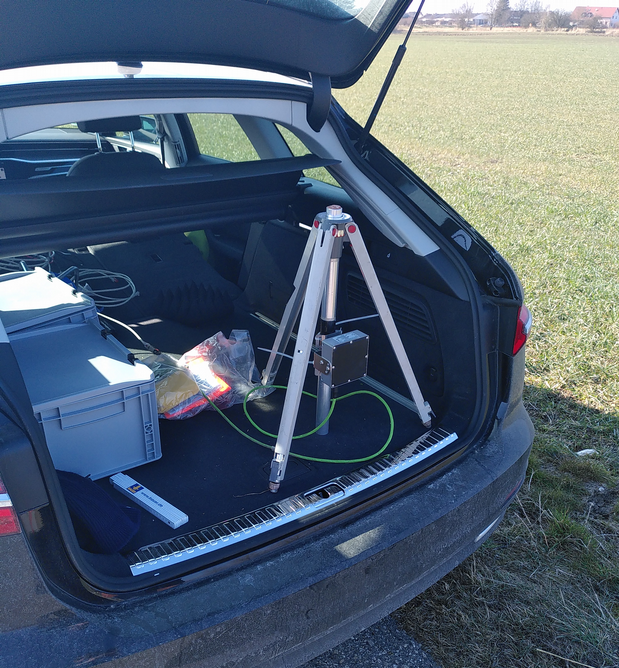}
		\vspace{-29ex}  
		\begin{flushleft}
			\hspace{-2ex}
			a)
		\end{flushleft}
		\vspace{29ex}  
	\end{minipage}\hfill
	\begin{minipage}{0.48\columnwidth}
		\hspace{0.2cm}
		\raggedright
		\includegraphics[trim={2.35cm 1.6cm 1cm 2cm},clip,angle=0, height = \columnwidth]{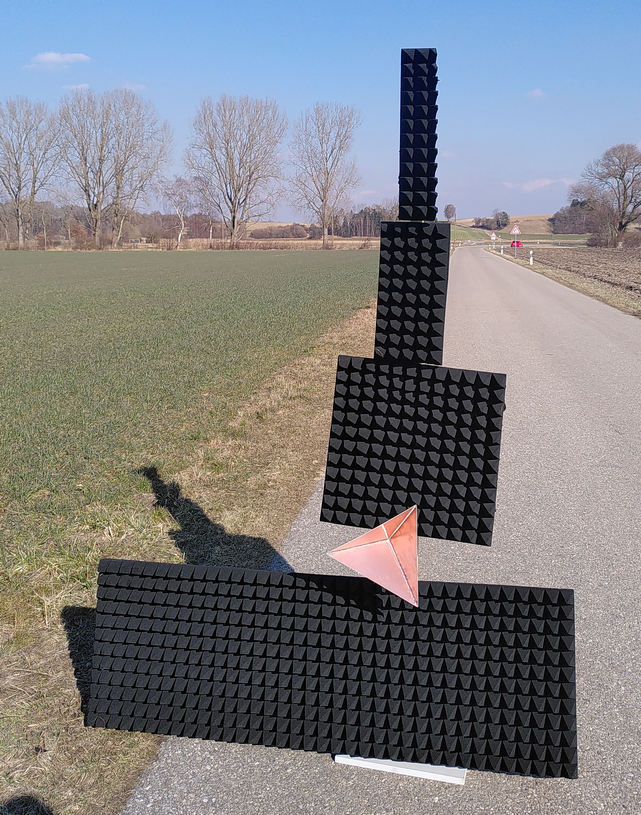}
		\vspace{-29ex}  
		\begin{flushleft}
			\hspace{0.3cm}
			\hspace{-4ex}
			b)
		\end{flushleft}
		\vspace{29ex}  
	\end{minipage}
	\vspace{-5ex}  
	\caption{Measurement setup: a) mobile sensor and b) stationary  reflector.}
	\label{fig:feat:hoehe:verif:SoRD:aufbau}
\end{figure}
\vspace{-0.2cm}
The trihedral corner reflector has a 29\,cm edge length and was installed on a pole in heights ranging from 0 to 2.5\,m.
The radar system was mounted at a constant height of 1.3\,m in the car, which moved along a straight line away from the reflector, doing repeated radar echo measurements.
A trihedral corner reflector was chosen, since it is a concentrated target with large RCS. A trihedral corner reflector only has a low bistatic RCS at larger angles between incoming and outgoing wavefront \cite{Zheng_2018_TripleRCS}. 
Referring to \Abb{fig:feat:hoehe:signalModel}, transmission from the direct to the indirect path (and vice versa) occurs only for large distances between the sensor and the corner reflector (such as those typical for S.\,o.\,MC.).
Smaller distances (such as those typical for S.\,o.\,PRI.) result in large angles between the paths and, thus, in a low bistatic RCSs. The echos that travel to the target on the direct path and back on the indirect path (or vice versa) do practically not exist, cf. \Gl{uebertragungsfunktionMitFp}.
Following the same procedure as above for the two-ray model, it can be shown that this results in the estimated heights for S. o. PRI. appearing at twice the reflector height. 
\subsection{Sampling over the measurement cycles (S.\,o.\,MC.)}
The spectra shown in \Abb{fig:feat:hoehe:test:erg:meascycl} were obtained using S.\,o.\,MC. Measurements with $T_{\mathrm{MC}} =$ 55.6\,ms at horizontal distances from 80 to 160\,m have been used to calculate the PSDs. A normalized PS as a function of the target height can be interpreted as a probability of the presence of a target at the corresponding height. 
The reflector heights are color-coded.
Three test runs with the vehicle moving at 2.8\,m/s were performed for each reflector height. The results are shown using solid, dashed, and dotted lines. The height estimates have an offset of about 20 to 30\,cm relative to the actual reflector height. 
However, for an application of the height estimate as a feature in a classification task, a monoton increasing height between the measurements is sufficient. 
\vspace{-0.2cm}
\begin{figure}[h!]
	\centering
	\includegraphics[trim={0.5cm 0cm 0cm 0.2cm},clip,width=\columnwidth]{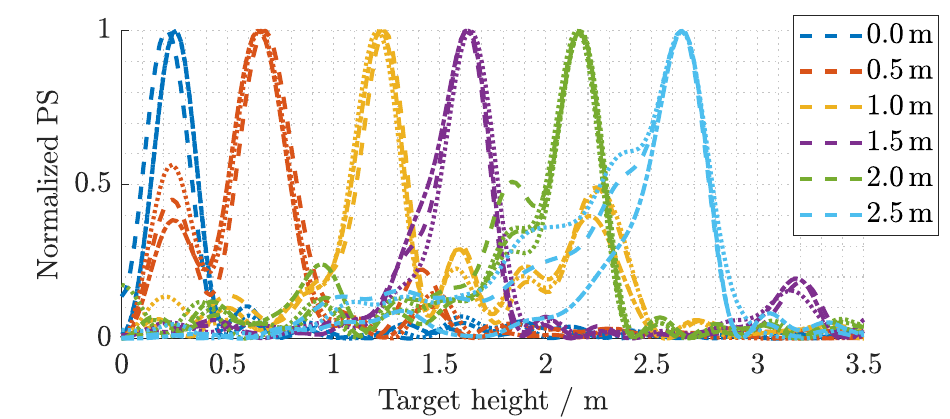}
	\caption{Spectra of all verification measurements scaled to the target height for S.\,o.\,MC. The reflector heights are color-coded. Three measurements were performed per target height. These are all drawn into the same figure.
	}
	\label{fig:feat:hoehe:test:erg:meascycl}
\end{figure}
\vspace{-0.3cm}
\subsection{Sampling over the ramp repetition interval (S.\,o.\,PRI.)}
\label{meas:PRI}
For the validation of S.\,o.\,PRI., the observation interval got limited to one CPI. That resulted in an observation interval (at 18\,m/s vehicle speed) of 71\,ms duration or 1.3\,m distance, that got sampled $\gls{numberoframps}=512$ times.
 This restriction allows an independent height estimation every measurement cycle.
In \Abb{fig:feat:hoehe:test:erg:distanceVsHeght}, each of the individual height estimates obtained with each of the measurement cycles is drawn as a dot as a function of the distance between the sensor and the vehicle at the corresponding measurement cycle.
Again, three test runs have been performed for each of the varying reflector heights. 
The method does not work for short distances between sensor and target since the maxima of the impulse responses of the echoes of the individual paths span across multiple radar range resolution cells there, cf. \Gl{eq::feat:hoehe:minDeltaR}.
The method has a defined limit to higher distances.
This limit is 0.66 times the resolution $\Delta h_T$ from \Gl{eq:feat:hoehe:delta_ht} and results from the DC removal on the propagation factor (\Abb{fig:feat:hoehe:test:erg:ablaufplan}). 
The estimated height is twice the reflector height because of the use of a trihedral corner reflector as a verification target, cf. Section \ref{sec:meas}.
\vspace{-0.2cm}
\begin{figure}[h!]
	\centering
	\includegraphics[trim={1cm 0cm 1cm 0.5cm},clip,width=\columnwidth]{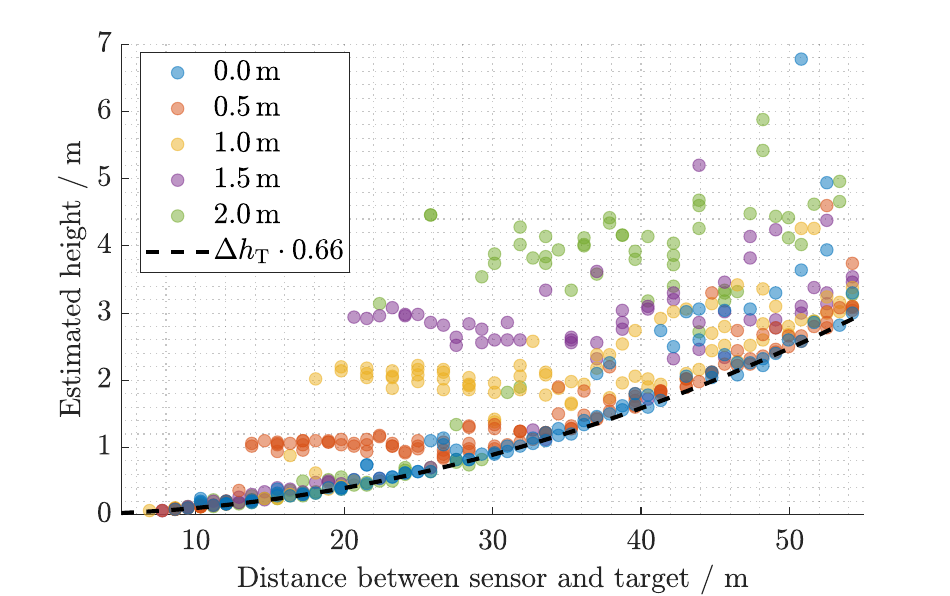}
	\caption{
	Maxima of the height spectra of the independent estimates at different distances between sensor and target for different target heights for S.\,o.\,PRI.}
	\label{fig:feat:hoehe:test:erg:distanceVsHeght}
\end{figure}
\vspace{-0.2cm}
\section{Conclusions}
The height estimation of a vehicle by analyzing the AM that originates from multipath propagation of the radar echoes seems feasible. Measurements with a trihedral corner reflector as a verification target show that the accuracy of the estimation depends, among other things, on the utilized sampling methods, which were chosen here with regard to
employing the concept in FMCW radar systems for automotive applications. As an outlook, the proposed method could also be adapted for estimating the heights of further road objects like street signs. The influence of other propagation paths as the ground, like guard rails, needs further investigation.

\bibliographystyle{IEEEtran}

\bibliography{IEEEabrv,IEEEexample,hoehe,grundlagen}

\end{document}